\documentclass[twocolumn,aps]{revtex4-1}

\usepackage{amsmath}
\usepackage{amssymb}
\usepackage{amsfonts}

\usepackage{graphicx}
\usepackage{framed}

\newcommand{\da}{^\dagger}

\newcommand{\br}[1]{\langle #1 \vert}
\newcommand{\ke}[1]{\vert #1  \rangle}
\newcommand{\bk}[2]{\langle #1  \vert #2  \rangle}

\newcommand{\xx}{\text{x}}  
\newcommand{\ir}{\text{s}}  
\newcommand{\pp}{\mathbf{p}}  
\newcommand{\kk}{\mathbf{k}}  
\newcommand{\Av}{\mathbf{A}}  

\begin{document}
\title{Monitoring Nonadiabatic Electron-Nuclear Dynamics in Molecules by Attosecond Streaking of Photoelectrons}

\author{Markus Kowalewski}
\email{mkowalew@uci.edu}
\author{Kochise Bennett}
\author{J\'er\'emy R. Rouxel}
\author{Shaul Mukamel}
\email{smukamel@uci.edu}
\affiliation{Department of Chemistry, University of California, Irvine,
California 92697-2025, USA}
\date{\today}%

\begin{abstract}
Streaking of photoelectrons has long been used for the temporal characterization of attosecond extreme ultraviolet pulses. When the time-resolved photoelectrons originate from a coherent superposition of electronic states, they carry an additional phase information, which can be retrieved by the streaking technique. In this contribution we extend the streaking formalism
to include coupled electron and nuclear dynamics in molecules as well as initial coherences
and demonstrate how it offers a novel tool to monitor non-adiabatic dynamics
as it occurs in the vicinity of conical intersections and avoided crossings. 
Streaking can enhance the time resolution and provide direct signatures of electronic coherences, which affect many primary photochemical and biological events.
\end{abstract}

\maketitle

\section{Introduction}
The rates and outcomes of virtually all photochemical and photobiological processes are dominated by conical intersections (CIs) \cite{Polli10,Rinaldi14,Barbatti:JPhotochemPhotobiol:2007,Sobolewski:PCCP:2002},
which provide a fast sub-100-femtosecond noradiative pathway back to the ground
state.
At a CI the electronic and nuclear degrees of freedom frequencies are comparable and strongly mix due to the breakdown of the Born-Oppenheimer approximation.
Available techniques for the detection of CIs include optical monitoring
of excited state populations \cite{Timmers14prl,mcfarland2014ultrafast,Polli10}, vibrational spectra \cite{Kowalewski15jcp,Fingerhut13JCPL,Oliver14,Raab98jcp},
electronic Raman techniques \cite{Kowalewski15prl,Hua16sd}
and photoelectron spectroscopy \cite{Horio09,Bennett16jctc,Trabs14jpb}.
Attosecond pulse sources
\cite{Helml14nphot,Pop10,grguravs2012ultrafast,bostedt2013ultra,
popmintchev2012bright,manzoni2014coherent} can directly access
the electron dynamics of molecular systems \cite{Alnaser14nc,Kraus15sci,Schultze14sci}.  This opens up the possibility of probing CIs by measuring the electronic coherences they generate.

Photoelectron spectroscopy (PES) \cite{Carlson75arpc} is a well established technique for exploring the electronic structure of molecules and solid-state systems \cite{PShufner}.
Its time-domain extension, time-resolved photoelectron spectroscopy (TRPES \cite{stolow2004femtosecond,Hudock07jpca}), is further capable of following the nuclear
dynamics on excited-state potential energy surfaces.
Signatures of electronic coherences and non-adiabatic dynamics can be detected through temporal oscillations on top of the conventional photoelectron signal \cite{Bennett16jctc,Trabs14jpb}. Typical energy gaps between molecular valence states
span a few eV range.
Sub-femtosecond pulses are thus necessary to resolve the beating pattern
in the time domain.
By exposing the generated photoelectrons to another strong infrared (IR) field, effectively modifying their kinetic energy, streaking can be used to detect their time of release from the
bound states \cite{Kienberger04nat,Avila06pra}. 
This method has been originally developed to characterize the shape of attosecond extreme ultraviolet (XUV) pulses \cite{Quere05jmo,Itatani02prl,Kitzler02prl}. 
The applied few-cycle IR pulse acts as a gate for the photoelectrons, commonly referred to as streaking.
It has been used to extract the quantum phase of the underlying wave function in atoms \cite{Yakovlev10prl}, the Wigner-Eisenbud time delay
in tunnel ionization processes \cite{Eckle08sci,Cirelli15ieee,Schultze10sci}, or the generation of
ultra-fast electron pulses \cite{Kirchner14np}.

In this paper, we
extend the theory of TREPS to include nonadibatic electron/nuclear dynamics in molecules. We demonstrate that by combining TRPES with the streaking technique, the phase of the molecular wavefunction may be recovered by the application of a few-cycle streaking field.
The patterns of electronic coherences already found in conventional, unstreaked TRPES can be enhanced by the streaking field, thus improving the temporal resolution. 
This can be used as an alternative to stimulated Raman techniques \cite{Kowalewski15prl}
recently proposed,
for the detection of molecular electronic coherences created at CIs. 
Unlike a conventional streaking setup, the system is prepared in a coherent superposition of electronic and nuclear degrees
of freedom which must be described in the joint electronic and nuclear phase space.
The pulse used to subsequently ionize the molecule covers at least half a cycle of the streaking field. 
This couples the momentum states in the free electron wave packet originating from different molecular states and introduces additional interference fringes attributable to electronic coherence. 

The paper is organized as follows. In the next section, we describe the formalism and demonstrate the spectral features along with simulations for purely electronic
atomic systems. Thereafter, we add the nuclear degrees of freedom and present simulations for a molecular model system undergoing nonadiabatic dynamics.

\section{Streaking; Theory and Schematics}
\begin{figure}
\centering
\includegraphics[width=0.4\textwidth]{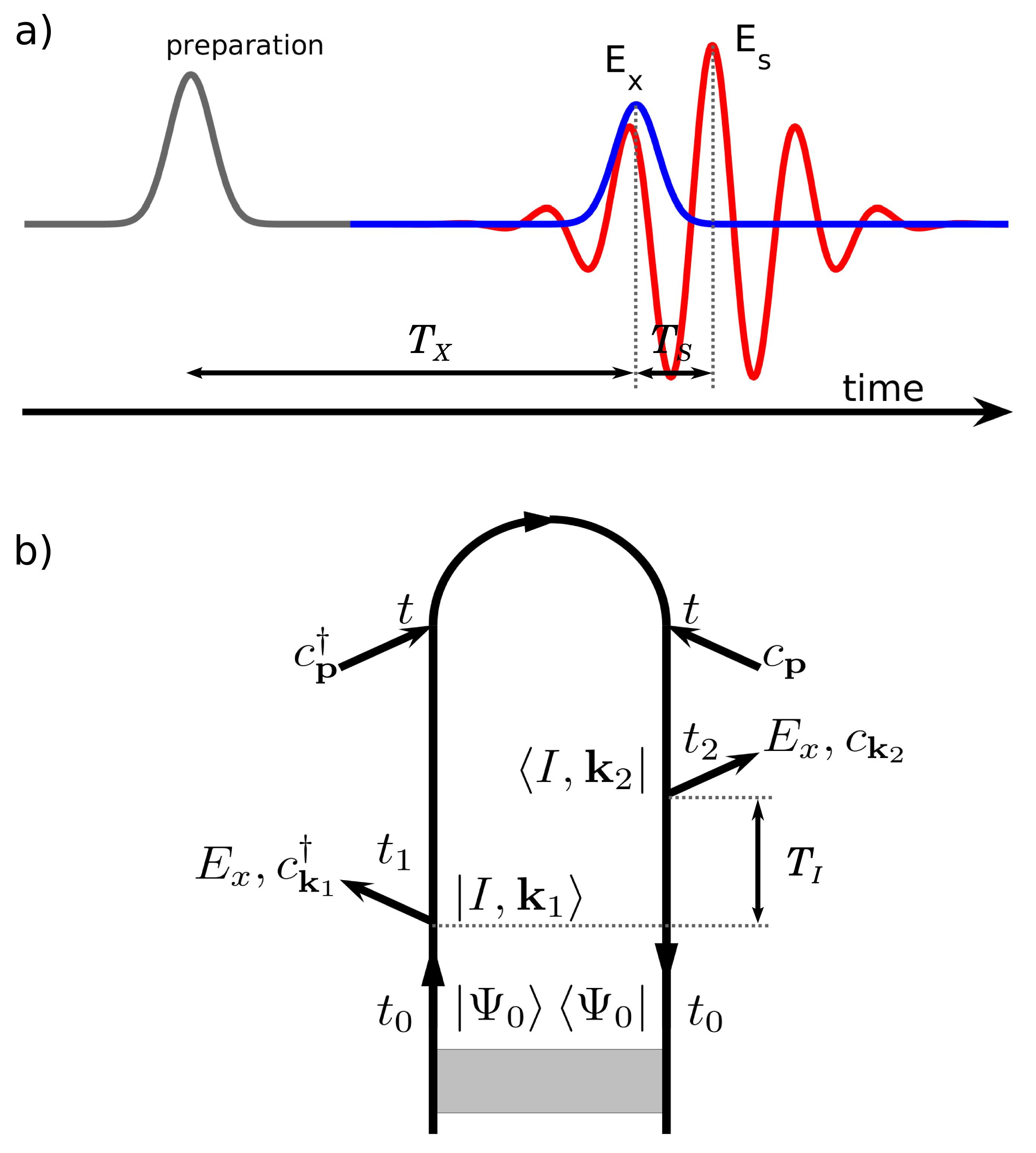}
\caption{(a) Pulse configuration for the streaked TRPES: An initial pulse prepares the molecule in a non-stationary superposition state
followed, after a delay $T_x$, by an ionizing pulse $E_x$ that has a temporal
overlap with the streaking field $E_s$.
 (b) Loop diagram for the streaking process (Eq. \ref{eq:strPESfinal}).
$\Psi_0$ represents the arbitrary molecular electron/nuclear wavepacket prepared by optical excitation.
The length of the time interval $T_I$ is determined by the matter evolution in Eq. \ref{eq:CM} and represents  the time it takes for Ionization to take place.
The photoelectrons are detected in momentum states $\ke{\pp}$ at time $t$ after the time evolution in the streaking field.}
\label{fig:diagram}
\end{figure}

Our derivation extends the perturbative description of TRPES in molecular systems \cite{Bennett16jctc} to take into account the free electron propagation under the influence of an IR streaking field as shown in Fig.\ \ref{fig:diagram}(a). 
An initial pump-pulse prepares the system in a non-stationary state, which is subjected to an ionizing pulse. 
The ionizing field $E_x$ and the streaking field $E_s$ overlap temporally.
The entire process is represented by the loop diagram \cite{Rahav10aamop} shown in Fig.\ \ref{fig:diagram}(b).

The system is described by the Hamiltonian
\begin{align}
H=H_{M}+H_x(t)+H_{es}(t) \,,
\end{align}
which consists of the molecular Hamiltonian with bound neutral and ionic states (indexed by $a$ and $I$ respectively)
\begin{align}
H_{M}=\hat T + \sum_{a}\hat{\varepsilon}_a\vert a\rangle\langle a\vert +\sum_{I}\hat{\varepsilon}_I\vert I\rangle\langle I\vert,
\end{align}
the minimal coupling Hamiltonian of the photoelectron in the presence of the streaking field
\begin{align}\label{eq:Hmc}
H_{es}(t)=\int \text{d}\mathbf{k}\big[ \mathbf{k}-\mathbf{A}(t)\big]^2\hat{c}_{\mathbf{k}}\hat{c}^\dagger_{\mathbf{k}},
\end{align}
and the interaction between ionizing x-ray pulse and the matter
\begin{align}
&H_{x}=-E_x(t)(\hat{\mu}+\hat{\mu}^\dagger)\\ \notag
&\hat{\mu}^\dagger=\int \text{d}\mathbf{k}\sum_{aI}\hat{c}_{\mathbf{k}}^\dagger\vert I\rangle\langle a\vert \hat{\mu}_{Ia}(\mathbf{k}).
\end{align}
where $\mu_{Ia}$ is the transition dipole moment between the neutral and ionic state,
$\hat T$ is the kinetic energy operator of the nuclei, and the potential energy operators of the molecular electronic states are given by $\hat{\varepsilon}_a$,
$\Av(t)=-\int_{-\infty}^t dt' \bold E_s(t')$ is the vector potential of the streaking field, and $\hat{c}_\kk\da$
is the creation operator of a photoelectron with kinetic momentum $\kk$.
Here, $\hat{c}_\kk\da$ are fermionic operators acting in the photoelectron space while $\hat\varepsilon_{a(I)}$ and $\hat{\mu}_{Ia}(\mathbf{k})$ are operators in the nuclear subspace.
\begin{figure*}
\centering
\includegraphics[width=1.0\textwidth]{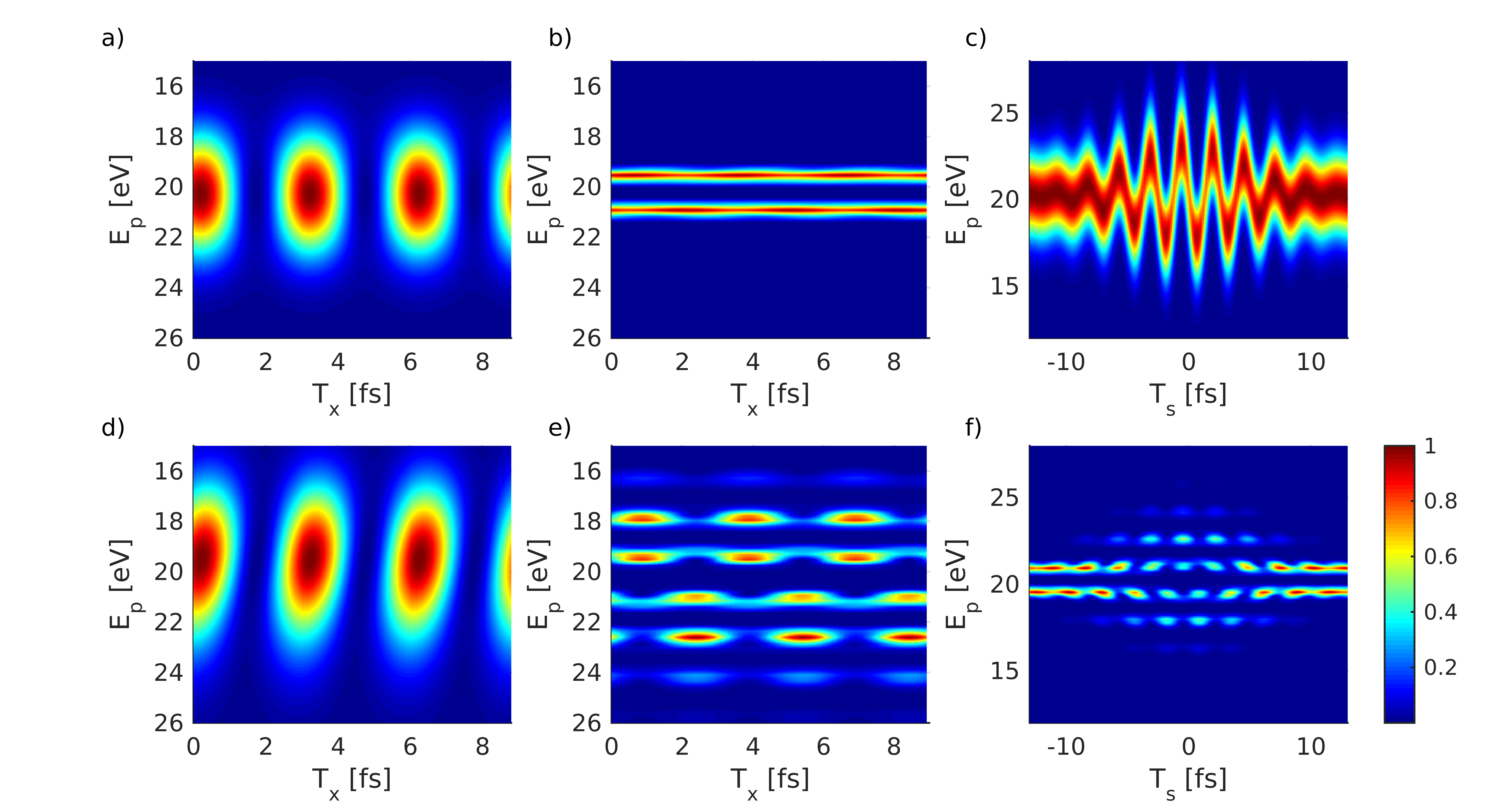}
\caption{Comparison of TRPES in a purely electronic (atomic) system for different pulse parameters:
(a) $\sigma_x$=0.5\,fs, the beating caused by the coherence is well resolved in the time domain. (b) $\sigma_x$=5.0\,fs, time resolution is lost. (d) like (a) but with the streaking
field applied (no major effect in this case). (e) like (b) but with the streaking field applied and the resolution is recovered.
(c) Streaking with $\sigma_x$=0.5\,fs recovers the structure of the streaking field
itself , while (f)  $\sigma_x$=5.0\,fs creates interference sidebands.}
\label{fig:TRPES}
\end{figure*}

The signal is given by the expectation value at the detection time $t$ of the photoelectron number operator $\hat N_\pp$.
This depends on the kinetic momentum $\pp$, the central time of the x-ray pulse $T_x$, and the streaking field parameters $\Lambda_s\equiv(T_s, \sigma_s,\omega_s,\phi_s)$
\begin{align}\label{eq:sdef}
S(\mathbf{p},T_x,\Lambda_s)=\langle \hat{N}_{\mathbf{p}}\rangle(t).
\end{align}
Expanding Eq. \ref{eq:sdef} to second order in $E_x$, as depicted in Fig. \ref{fig:diagram}(b), yields or key result:
\begin{widetext}
\begin{align}\label{eq:strPESfinal}
S_\text{e}(\pp,T_\xx,T_\ir)\approx \int dt_1\int dt_2 &\tilde{E}^*_{\xx}(t_1-T_x)\tilde{E}_{\xx}(t_2-T_x)
 C_M(t_2,t_1) e^{-i(\frac{\pp^2}{2}-\omega_\xx) (t_2-t_1)}\\\nonumber
 &\times\exp \left[ -i \pp\int_{t_1}^{t_2} d\tau \tilde{\Av}(\tau+T_s-T_x) \right]
\end{align}
\begin{equation}\label{eq:CM}
C_M(t_2,t_1)=\br{\Psi_0} U_M\da(t_2,t_0)  \mu(\pp-\Av(t_2)) U_M\da(t_2,t_1) \mu(\pp-\Av(t_1)) U_M(t_1,t_0) \ke{\Psi_0}
\end{equation}
\end{widetext}
Here the effects of the molecular bound states are contained in the correlation function $C_M$ (note that we use a tilde to indicate shifted field envelopes centered at zero argument).
The molecular propagator $U_M$ depends on  the full nuclear+electronic molecular Hamiltonian including non-adiabatic couplings.
We have assumed that the photoelectron wavepacket does not interact with the molecular ion, which is reasonable for sufficiently fast photoelectrons \cite{Zhang08prl}.
The streaking field must be weak enough to not perturb the molecular eigenstates or ionize
the molecule, which also justifies dropping the $\Av^2$ interaction in the minimal coupling Hamiltonian (Eq. \ref{eq:Hmc}), as done in Eq. \ref{eq:strPESfinal}.
The initial state $\ke{\Psi_0}$ is given by a product of a non-stationary molecular state and the photoelectron vacuum state. 
When the nuclear degrees of freedom are neglected, Eq. \ref{eq:strPESfinal} reduces to the
modulus squared of an amplitude \cite{Yakovlev10prl} and can be recast as a FROG trace, which allows the inversion of signals to yield the ionization pulse shape or the quantum phase of the atomic state (for a detailed derivation see SI).
However, more generally, the nuclear and electronic degrees of freedom are coupled and the matter correlation function $C_M$
depends in a non-trivial way on the propagation of the molecular wave packet
in the interval $T_I=[t_1,t_2]$ where the ionization takes place (see Fig. \ref{fig:diagram}).
The precise definition of the ionization time and its detection has drawn considerable attention \cite{Eckle08sci,Cirelli15ieee,Schultze10sci}.
The time-dependence, via the nuclei, of the electronic eigenvalues 
is the main difference from atomic
experiments.
In the limit of an impulsive ionization event $T_I\rightarrow0$ and Eq. \ref{eq:strPESfinal} captures 
a snapshot of the system that depends on $T_x$ and $T_s$.
When the ionization event duration $T_I$ is long enough to allow for nuclear motion,
the signal also depends on the time evolution during the $T_I$ interval 
where there is a coherence between the electronic states of the neutral molecule (see Fig. \ref{fig:diagram}(b)).
The evolution during the coherence period is given by Eq. \ref{eq:CM}.
The propagator $U\da_M(t_1,t_2)$ describes the evolution of the nuclear wave packet on the
potential energy surface of the ionic state.

To set the stage we first present the basic features of the streaked photoelectron spectra with initial coherence for a purely electronic atomic model system in order to illustrate the
signatures of purely electronic coherence in the streaking signal.
The molecular system with the coupled electron and nuclear dynamics is discussed in next section.
We assume a two level atom with bound states $\vert g\rangle$ and $\vert e\rangle$, $\omega_e-\omega_g =  1.36$\,eV, and a single ion state with ionization energies $(\omega_{Ig}, \omega_{Ie}) =   (5.44,4.08)$\,eV. 
The oscillation period due to the coherence in the streaking field-free photoelectron spectrum is $\approx 3$\,fs, setting an upper bound to the ionization pulse length for the observation of the beating pattern in a conventional TRPES experiment.
The system is prepared initially in a coherent superpostion of $\vert g\rangle$ and $\vert e\rangle$ and is subsequently ionized by a XUV Gaussian pulse of $\omega_x=25$\,eV central frequency and $\sigma_x$ (full width at half maximum, FWHM) duration:
\begin{align}
\tilde E_x(t) = E_x e^{-t^2/0.72\sigma_x^2}
\end{align}
The streaking field vector potential is given by
\begin{align}
\tilde{\Av}(t) = \bold E_s \int_{\infty}^{t} \mathrm{d}\tau \cos(\omega_s \tau+\phi_s) e^{-\tau^2/0.72\sigma_s^2}\,,
\end{align}
where $\omega_s$ is its carrier frequency and $\phi_s$ the carrier envelope phase.
Figures \ref{fig:TRPES}(a) and \ref{fig:TRPES}(b) show the bare, unstreaked photoelectron spectra for two ionization pulse lengths.
The FWHM of the ionization pulse used in Fig.\ \ref{fig:TRPES}(a) is 0.5\,fs and
can temporally resolve the beating pattern.
However, the states $\ke{g}$ and $\ke{e}$ can not be distinguished along the $E_p$ axis due to the broad width of $E_x$. 
The longer ionization pulse 5\,fs used in Fig. \ref{fig:TRPES}(b) allows
for a clear resolution of states $\ke{g}$ and $\ke{e}$ in the frequency domain but is too long to resolve their time domain
beating pattern.
Figures \ref{fig:TRPES}(d) and (e) show the photoelectron spectra under the influence of a streaking field ($T_s=0$, $\sigma_s=8$\,fs, $\omega_s=1.6$\,eV, $\phi_s=0$) for $\sigma_x=0.5$\,fs and $\sigma_x=5$\,fs respectively.
The beating pattern is recovered by the spreading of the photoelectron peaks in Fig. \ref{fig:TRPES}(e), where the side bands are generated at $E_p=\varepsilon_{a/b}\pm n \omega_s$ for integer $n\geq1$.
Typical streaking spectra are shown in Figs.\ \ref{fig:TRPES}(c) and (f). With an ionization pulse length
shorter than the optical cycle of the streaking field (Fig.\ \ref{fig:TRPES}(c)) the pattern of the streaking field is recovered. However, the two states may not be resolved by the photoelectron energy.
For an ionization pulse length covering a full optical cycle of the streaking field (Fig.\ \ref{fig:TRPES}(f)) the frequency resolution is retained. The pattern in
the photoelectron kinetic energy $E_p$ is generated
by the side bands in ($E_p=\varepsilon_{a/b}\pm n \omega_s$), while the oscillatory pattern in $T_s$ is a clear
signature of the coherence.
An eigenstate or an incoherent mixture of two states would not show the beating pattern, i.e., we would see straight lines along $T_x$.

\begin{figure}
\centering
\includegraphics[width=0.5\textwidth]{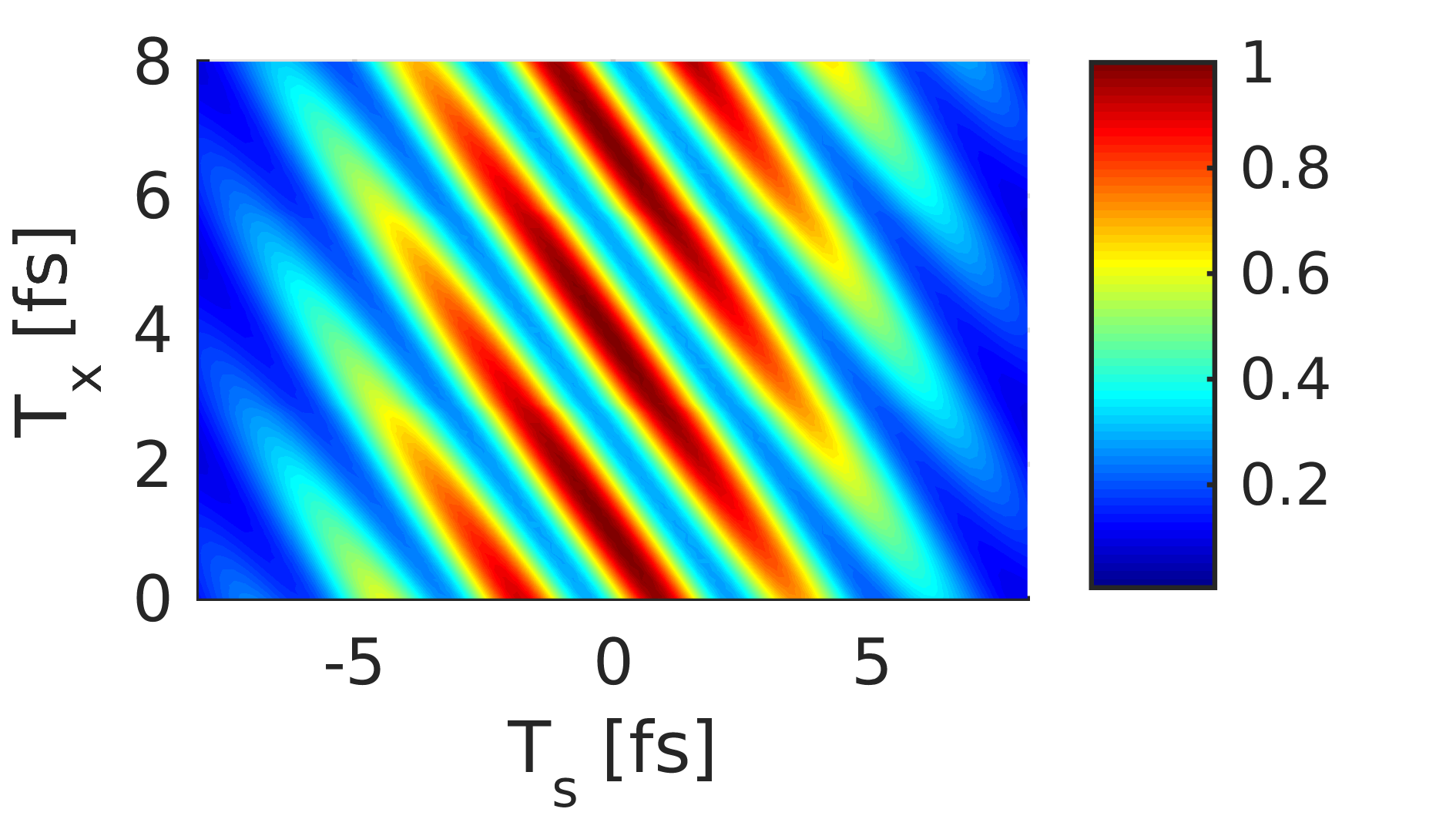}
\caption{Streaking signal in purely electronic atoms vs.\ the
ionization and streaking pulse delay. The diagonal pattern is caused
by the coherences.}
\label{fig:TsTx}
\end{figure}
Figure \ref{fig:TsTx}
shows the signal vs. $T_x$ and $T_s$ at fixed $E_p=$18\,eV.
This representation can be used as an indicator for the initial coherence:
The diagonal pattern is caused by the quantum phase of the superposition (i.e., the time dependence of the beating pattern).
In case of incoherent states the signal is independent of $T_x$ yielding a purely vertical pattern.
This clearly shows the capability of regaining time resolution when the streaking side bands
of the two states coincide ($\varepsilon_{a}\pm n_a \omega_s=\varepsilon_{b}\pm n_b \omega_s$).

\section{Streaking Detection of Nonadiabatic Dynamics}
\begin{figure}
\centering
\includegraphics[width=0.45\textwidth]{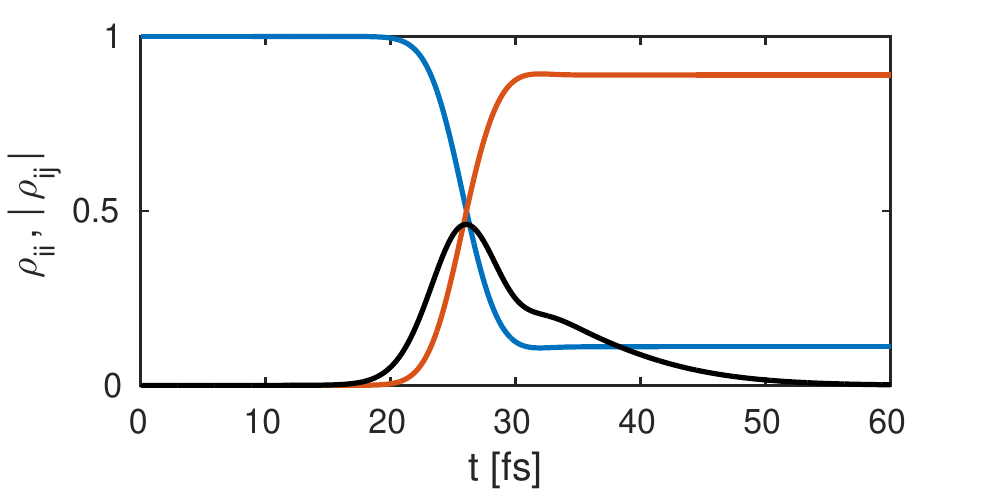}
\caption{Time evolution of the electronic populations (state $a$ blue and state $b$ red) for the 1-dimensional molecular model.
The magnitude of the electronic coherence ($\rho_{ab}=\bk{\phi_a}{\phi_b}$) is indicated by the black curve.
At $\approx 20$\,fs the molecule hits the avoided crossing.}
\label{fig:pops_1D}
\end{figure}
We now demonstrate the signatures of nonadiabatic
dynamics in the signal using a harmonic model with a single vibrational mode.
The model has two electronic states, represented
by two symmetrically displaced harmonic oscillators, a Gaussian diabatic coupling, and
a harmonic ion state (for details of the model see SI). The quantum dynamics simulation starts out with a displaced Gaussian
wave packet as its initial condition and hits the curve crossing at $\approx 20$\,fs, creating
an electronic coherence (see Fig.\ \ref{fig:pops_1D}).
This simple model can be solved exactly using a numerical grid (see SI).

\begin{figure*}
\centering
\includegraphics[width=1.0\textwidth]{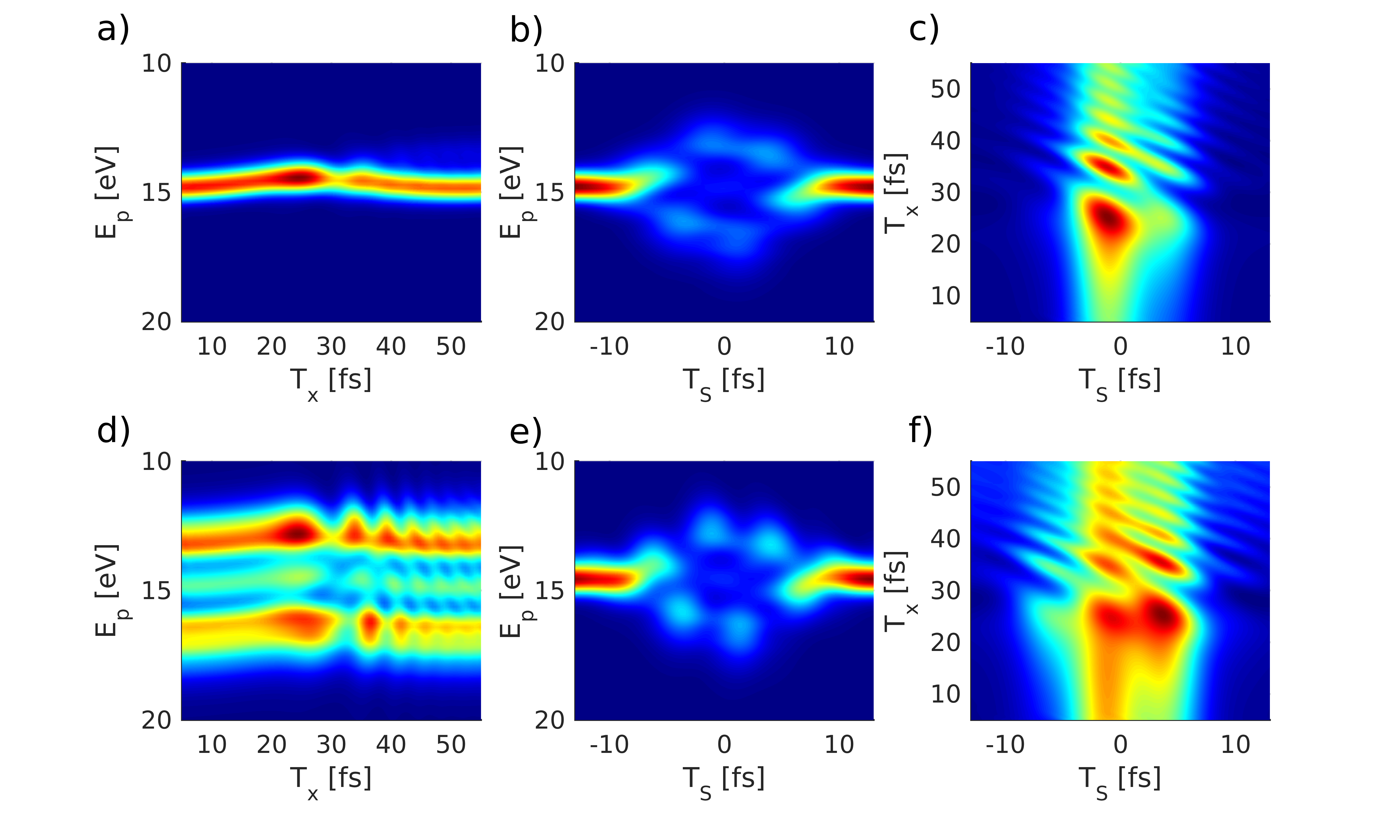}
\caption{Streaking spectra for the molecular 1D model with parameters $(\sigma_s,\sigma_x) = (8.0,5.0)$\,fs, $(\omega_s,\omega_x)=(0.82,20)$\,eV. TRPES without (a) and with (d) streaking field ($T_s=0$).  Photoelectron energy vs.\ streaking delay for $T_x=7$\,fs (b) $T_x=35$\,fs (e).
Streaking delay vs. ionization delay for  $E_p=12$\,eV (c) and $E_p=13$\,eV (f).}
\label{fig:streaking1D}
\end{figure*}
A set of streaking spectra resulting from the dynamics are shown
in Fig. \ref{fig:streaking1D}.
The ionization pulse length used (5\,fs FWHM) is not capable of resolving
the coherent beating pattern in a bare PES (Fig. \ref{fig:streaking1D}(a)).
However, the application of the streaking pulse (Fig.\ \ref{fig:streaking1D}(d)) shows a clear signature of coherent oscillations for $T_x>20$\,fs (i.e., after the molecule has reached
the avoided crossing). The PES is stretched along $E_p$ thus effectively increasing
the time resolution by distributing the photoelectrons over $E_p$ according to their release time.
The typical streaking representation ($E_p$ vs.\ $T_s$) is shown in Figs.\ \ref{fig:streaking1D}(b) and (e) for different ionization delays $T_x$. The pattern of the streaking
pulse is blurred since the ionization pulse covers more than a full cycle
of the streaking field.
The representation in Figs.\ \ref{fig:streaking1D}(c) and \ref{fig:streaking1D}(f)
makes it clear that streaking puts a time stamp on the photoelectrons.
The beating pattern along $T_x$ appears at around 30\,fs and creates lines on
the diagonal $T_x$/$T_s$ as clear indicator of the electronic coherence created by
the avoided crossing. Unlike Fig. \ref{fig:TsTx}, the pattern is not symmetric
with respect to $T_s$. This is due to the nuclear motion: electrons released at different
times originate from different nuclear configurations.

\section{Conclusions}
We have extended the description of TRPES in molecules to account for
the effect of an IR streaking field.
This strong field couples the momentum states of the free
electron wavepacket and thus allows the recovery of the electronic coherences
imprinted in it with a higher resolution than that of bare TRPES.
The streaking field clocks the photolectron release time
by spreading them over a range of kinetic energies.
These features are simple for an atomic system and are routinely used to
characterize attosecond pulses for a given matter dynamics. Here, we
demonstrate that a reverse objective can be met, i.e. measurement
of the matter dynamics knowing the pulses, and that it can be extended
to molecular systems with strongly coupled electron and nuclear dynamics.

From our model calculations, it becomes clear that the
streaking field could be used to detect avoided crossings and
conical intersections in molecules. In the presence of nuclear dynamics,
the signal may not longer be recast as an amplitude squared since the
wavepacket evolves non-trivially between the two interactions with the
ionization x-ray pulse. This evolution is responsible for the loss of
symmetry along $T_s$ in the streaking spectra and can be used to infer
and quantify the underlying nuclear dynamics.

\begin{acknowledgements}
The support of the Chemical Sciences, Geosciences, and Biosciences division, Office of Basic Energy Sciences, Office of Science, U.S. Department of Energy through Award No. DE-FG02-04ER15571 and of the National Science Foundation (Grant No CHE-1361516) is gratefully acknowledged. K.B. was supported by the DOE award.
M.K gratefully acknowledges support from the Alexander von Humboldt foundation
through the Feodor Lynen program.
\end{acknowledgements}

\end{document}